\documentclass[useAMS,usenatbib]{mn2e} 
\usepackage{graphicx}

\def\a{{$\alpha$}}

\def \kms{\ifmmode{~{\rm km\,s}^{-1}}\else{~km~s$^{-1}$}\fi}  
\def \vhel{\ifmmode{V_{{\rm hel}}}\else{$V_{{\rm hel}}$}\fi}  
\def \vsys{\ifmmode{V_{{\rm sys}}}\else{$V_{{\rm sys}}$}\fi}  
\def \vlsr{\ifmmode{V_{{\rm lsr}}}\else{$V_{{\rm lsr}}$}\fi}  
\def \vobs{\ifmmode{V_{{\rm obs}}}\else{$V_{{\rm obs}}$}\fi}  
\def \degree{\ifmmode{^{\circ}}\else{$^{\circ}$}\fi}  
\def \lsun{\ifmmode{{\rm\ L}_\odot}\else{${\rm\ L}_\odot $}\fi}  
\def \msun{\ifmmode{{\rm\ M}_\odot}\else{${\rm\ M}_\odot$}\fi}  
\def \myr{\ifmmode{{\rm\ M}_\odot{\rm\ yr}^{-1}}\else{${\rm\ M}_\odot$   
yr$^{-1}$}\fi}  
\def \teff{\ifmmode{{\rm{T}}_{\rm eff}}\else{${\rm{T}}_{\rm eff}$}\fi}  
\def \mdot{\ifmmode{{\rm\dot{M}}}\else{${\rm\dot{M}}$}\fi}  
  
\newcommand{\h}{$^{\rm h}$}  
\newcommand{\m}{$^{\rm m}$}  
\newcommand{\s}{$^{\rm s}$}  
\newcommand{\dd}{$\delta$}  
\newcommand{\ha}{H$\alpha$}  
\newcommand{\hbeta}{H$\beta$}  
  
\newcommand{\hi}{H{\sc i}}  
\newcommand{\HNII}{{\rm H}$\alpha+$[N {\sc ii}]~6548~\&~6584~\AA}  
\newcommand{\hnii}{{\rm H}$\alpha+$[N {\sc ii}]}  
\newcommand{\nii}{[N~{\sc ii}]~6584~\AA}  
\newcommand{\oi}{[O~{\sc i}]}

\newcommand{\oiii}{[O~{\sc iii}]~5007~\AA}  
\newcommand{\sii}{[S~{\sc ii}]~6717 \& 6731~\AA}  

\newcommand{\nitrogen}{[N~{\sc ii}]}  
\newcommand{\oxygen}{[O~{\sc iii}]}  
\newcommand{\sulfur}{[S~{\sc ii}]}  
\newcommand{\vel}{\rm km s$^{-1}$}  
\newcommand{\flux}{$10^{-17}$ erg s$^{-1}$ cm$^{-2}$ arcsec$^{-2}$}  
\newcommand{\siirat}{[S~{\sc ii}]}  
\newcommand{\niiab}{[N\,{\sc ii}]\ 6548,\ 6584\,\AA} 
  
\title[Deep optical observations of  
the W50--SS 433 system.]{Deep optical observations of the interaction of the 
SS 433 microquasar jet with the W~50 radio continuum shell.}  
\author[P. Boumis et al.]{P. Boumis$^{1}$\thanks{E-mail:
ptb@astro.noa.gr}, J. Meaburn$^{1,2}$, J. Alikakos$^{1,3}$,
M. P. Redman$^{4}$, S. Akras$^{1,3}$, \newauthor F. Mavromatakis$^{5}$, 
J. A. L\'{o}pez$^{6}$, A. Caulet$^{6}$  and C. D. Goudis$^{1,3}$\\
$^{1}$Institute of Astronomy \& Astrophysics, National Observatory of  
Athens, I. Metaxa \& V. Paulou, GR--152 36 P. Penteli, Athens,  
Greece.\\  
$^{2}$Jodrell Bank Observatory, University of Manchester,  
Macclesfield SK11 9DL, UK.\\  
$^{3}$Astronomical Laboratory, Department of Physics, University  
of Patras, 26500 Rio--Patras, Greece.\\  
$^{4}$Department of Physics, National University of Ireland Galway,  
Galway, Ireland.\\  
$^{5}$Technological Education Institute of Crete, General Department of  
Applied  
Science, P.O. Box 1939, GR-710 04 Heraklion, Crete, Greece.\\  
$^{6}$Instituto de Astronomia, UNAM, Apdo. Postal 877. Ensenada, B.C. 22800,  
M\'{e}xico.\\ 
}  
  
\begin{document}  
  
\date{Accepted 2007 July 24. Received 2007 July 24; in original form 2007 March 30}  
  
\pagerange{\pageref{firstpage}--\pageref{lastpage}} \pubyear{2007}  
  
\maketitle  
  
\label{firstpage}  
  
\begin{abstract}  
  
Four mosaics of deep, continuum--subtracted, CCD images have been
obtained over the extensive galactic radio continuum shell, W~50,
which surrounds the remarkable stellar system SS~433. Two of these
mosaics in the \hnii\ and \oiii\ emission lines respectively cover a
field of $\sim 2\degr.3 \times 2\degr.5$ which contains all of W~50
but at a low angular resolution of 5 arcsec. The third and fourth
mosaics cover the eastern (in \oiii) and western (in \HNII)
filamentary nebulosity respectively but at an angular resolution of
1 arcsec.  These observations are supplemented by new low dispersion
spectra and longslit, spatially resolved echelle spectra.  The \oiii\
images show for the first time the distribution of this emission in
both the eastern and western filaments while new \hnii\ emission
features are also found in both of these regions.
Approaching flows of faintly emitting material 
from the bright eastern filaments of up 100~\kms\ in radial velocity
are detected. The present observations also suggest that the heliocentric
systemic radial velocity of the whole system is 56 $\pm$~2~\kms.
Furthermore, very deep imagery and high resolution spectroscopy of a
small part of the northern radio ridge of W~50 has revealed for the
first time the very faint optical nebulosity associated with this
edge. It is suggested that patchy foreground dust 
along the $\approx$ 5~kpc sightline 
is inhibiting the
detection of all of the optical nebulosity associated with W~50.  
The interaction of the microquasar jets of SS~433 with the W~50 
shell is discussed.

\end{abstract}  
  
  
\section{Introduction}  


Most of the galactic supernova remnants (SNRs) have been identified by
their radio, optical and X--ray emission \citep{gre06}. G 39.7$-$2.0
(W~50) was identified in radio wavelengths by \citet{wes58} and first
classified as an evolved SNR by \citet{hol69}. Its radio continuum
image has a complex `seashell' appearance showing a main spherical
shell of non--thermal emission ($\sim$58 arcmin in diameter) with
easterly and westerly extensions or lobes. The radio spectral index
varies across the eastern, central and western areas between values of
0.4 to 0.7 and the total flux densities are 71 Jy and 160 Jy at 1465
and 327 GHz, respectively \citep{dub98}. The latter authors also
presented the latest \hi\ observations showing some evidence of an
interaction between W~50 and its surrounding interstellar medium (ISM)
at a velocity $\sim$76 \vel\ while a distance to W~50 of $\sim$3 kpc
was also calculated. This simple SNR interpretation of W~50 is
complicated by the presence at its centre of the remarkable stellar
binary system SS~433 (see \citealt{fab04} for a comprehensive review)
from which emission is ejected in two oppositely directed relativistic
jets aligned along the same axis as the elongation of the radio lobes
(see also fig.2; \citealt{dub02}). The two jets are also X--ray
emitters (\citealt{bri96}; \citealt{saf97}) showing different
morphological and spectral properties. These X--ray and IRAS
\citep{ban87} observations suggest that the interaction of SS~433 jets
with the surrounding ISM plays an important role in the shaping of the
envelope. \citet{sti02} measured the distance to the jet of
SS~433, hence W~50, from radio proper--motion measurements, as 4.61
$\pm$ 0.35 kpc and Blundell \& Bowler (2004) give 5.5 $\pm$ 0.2 kpc.
Here we will adopt a distance of $\approx$ 5 kpc.
  
Optical emission was first discovered by both \citet{van80} and
\citet{zea80} who found two groups of faint filaments, one east and one  
west of SS~433 at a distance of $\sim$30 arcmin from
it. Spectroscopic observations (\citealt{kir80}; \citealt{mur80};
\citealt{shu80}) show that all the optical filaments originated from
shock--heated gas since the \sulfur/\ha$\sim$2 while very strong
\nitrogen\ (\nii/\ha$\sim$3) emission is also present.  Faint \oiii\
emission was also detected in these spectra (\oiii/\ha$\sim$0.4) but
only from the eastern filaments. \citet{maz83} presented high spectral
but low angular (with 120 arcsec \& 200 arcsec beam diameters)
resolution profiles of the \nii\ line from both the eastern and
western optical filaments of W~50. These profiles have a complex
structure and are broad ($\sim$50 \vel) and see Sect. 3.3 for further
details and comparison with new line profiles.
  
A large number of papers have been published (see the review of
\citealt{fab04} and references therein) concerning the association of
W~50 with SS~433. A consensus is emerging that the SS~433 jets have
punched holes in a preceding, expanding supernova shell
(e.g. \citealt{vel00}) though the formation of the whole structure of
W~50 by the jets alone has not been completely ruled out (\citealt{fab04}).
  
In the present paper, the first continuum--subtracted, deep CCD images
of the total area of W~50 in the light of \hnii\ and \oiii\ are
presented in a mosaic which covers an area of $\sim 2\degr.3 \times
2\degr.5$. The major eastern and western regions of nebulosity are
also imaged at higher angular resolution in the \oiii\ and \HNII\
lines respectively to reveal the fine structure of the optical
filaments. Furthermore, a very deep \HNII\ image of a small region of
the northern ridge of radio emission has been obtained in an attempt
to detect hitherto undiscovered filamentary nebulosity that would be
expected if the W 50 radio shell is a supernova remnant.
  
The new, faint, optical emission found over W~50, together, with its
comparison with a high resolution radio map, permits a better
appreciation of the morphology of W~50 and its interaction with the
jets of SS~433. In addition, new, deep, low--resolution and
spatially--resolved, high--resolution spectra of this optical
nebulosity have been obtained to advance the understanding of this
interaction.
  
\section[]{Observations \& Results}  

A summary and log of our imaging and spectral observations is provided
in Table 1. In the sections below, we describe the details of these
observations.
  
\subsection{Imaging}  
  
\subsubsection{Wide field imagery}  
  
The wide--field imagery was taken with the 0.3 m Schmidt--Cassegrain
(f/3.2) telescope at Skinakas Observatory in Crete, Greece from 27 to
30 June, 2003.  A 1024 $\times$ 1024 SITe CCD was used which has a
pixel size of 24 $\mu$m resulting in a 89 $\times$ 89 arcmin$^{2}$
field of view and an image scale of 5 arcsec per pixel.
  
Each of four different fields was observed for 2400 s in both filters
while corresponding continuum images were also observed (180 s each)
and were subtracted from those containing the emission lines to
eliminate the confusing star field (see
\citealt{bou02} for details of this technique). All fields were  
projected on to a common origin on the sky and were subsequently
combined to create the final mosaics in \hnii\ and \oiii. During the
observations the `seeing' varied between 0.8 to 1.2 arcsec. The
image reduction was carried out using the IRAF and STARLINK
packages. The astrometric solutions were calculated for each field
using reference stars from the Hubble Space Telescope (HST) Guide Star
Catalogue \citep{las99}. All coordinates quoted in this paper refer to
epoch 2000.

The images of the nebulosity in all of the fields considered here are
detected, at most, only a few times greater than the residual noise
level. In these circumstances we have always chosen to display the
data with a linear scale, but negatively, and at high contrast. Also
we have chosen not to suppress the noise artificially, e.g. by
excessive smoothing or lifting the zero level, for this often leads to
detection artefacts being confused in the resultant display with real
nebulous features.
  
\begin{figure*}  
\centering  
\scalebox{0.90}{\includegraphics{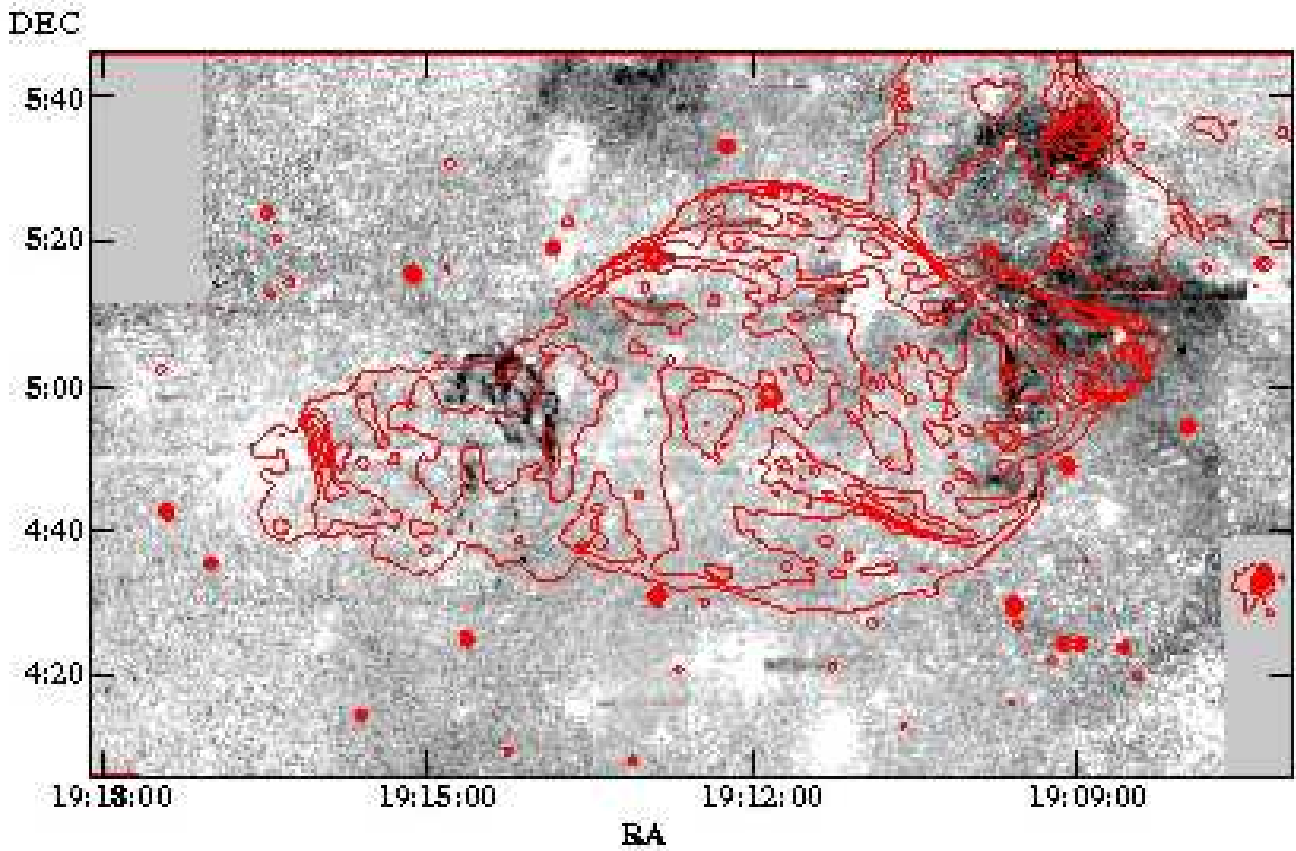}}  
\caption[]{The correlation between the $\sim 2\degr.3 \times
2\degr.5$~negative continuum--subtracted mosaic of W50 in the light of
\hnii\ and the radio emission at 1465 MHz (solid lines). The 1465 MHz
\citep{dub98} radio contours scale linearly from $1 \times
10^{-2}$~Jy/beam to 0.1 Jy/beam. The strong radio source to the
north--west is LBN 109 (see text).The image has been smoothed to
suppress the residual from the imperfect continuum subtraction. The
horizontal line segments seen near overexposed stars in this figure
and the next figures are due to the blooming effect. The optical
features are shown in detail in Figs. 2(b) and 3(b).}
\label{fig01}  
\end{figure*}

The image of W~50 with this system is shown in Fig. \ref{fig01}. This
is a mosaic of four images taken through the \hnii\ filter. The same
field was observed with the \oiii\ emission line filter.  
\citet{dub02} and \citet{dub98} using their high resolution radio map  
of W~50 made a comparison between the radio/X--ray/\hi\ emission.
These radio continuum contours (\citealt{dub98}) are compared in Fig. 1
with this new mosaic of \hnii\ images.
  
Selected areas of the eastern and western filaments seen in Fig. 1 are
shown respectively in Figs. 2a \& b and 3a \& b in the light of \oiii\
and \hnii. For the first time, deep, continuum--subtracted, CCD images
in the light of \hnii\ and \oiii\ of W~50 have been obtained. The
\hnii\ mosaic of images in Fig. 1 shows new filamentary and diffuse
emission while, for the first time, \oiii\ filamentary emission from
W~50 is revealed.  The most striking features are the differences in
the filamentary nebulosities in Figs. 2a \& b. The \oiii\ emission in
Fig. 2a forms a 24 arcmin long outer arc as far as $\alpha \simeq$
19\h14\m30\s, $\delta \simeq$ 4\degr30\arcmin\ which contains the
predominantly \nii\ (see Sect. 2.2.1) emitting filaments. This eastern
filamentary arc is convex with respect to SS~433 whereas the \nii\
emitting western filamentary arc in Figs. 3a \& b is concave and only
has very localised \oiii\ emitting counterparts.

Diffuse \hnii\ is present to the north--east (\a\ $\simeq$
19\h13\m05\s, \dd\ $\simeq$ 5\degr08\arcmin00\arcsec) and north--west
(\a\ $\simeq$ 19\h11\m30\s, \dd\ $\simeq$ 5\degr12\arcmin00\arcsec)of
the main, circular radio remnant of W~50 in Fig. 1. It is also present
within its western radio lobe and the possibility that it belongs to
the remnant cannot be ruled out.  The weak diffuse emission which is
present north of W~50 (\a\ $\simeq$ 19\h13\m05\s, \dd\ $\simeq$
5\degr43\arcmin00\arcsec) and to the north--west (the bright, extended
nebula LBN 109; \citealt{lyn65}) is outside W~50's radio borders and
not associated with the remnant.
  
\subsubsection{High resolution imagery of the eastern and western filaments}  
  
Optical images at higher angular resolution of the eastern and western
filaments were also obtained with the 1.3 m (f/7.7) Ritchey-Cretien
telescope at Skinakas Observatory during 2005 in September 5, 9--10
and 2006 in July 27--28 using an \oiii\ and an \hnii\ interference
filters, respectively. The detector was a 1024$\times$1024 SITe CCD
with a field of view of 8.5 $\times$~8.5 arcmin$^{2}$.  Ten exposures
through the \oiii\ filter each of 2400 s duration and ten
corresponding exposures in the continuum, each of 180 s, were taken of
the eastern filaments, and similarly two \HNII\ and continuum images
of the western filaments were obtained. The continuum subtracted
mosaics of these images are shown in Fig. 4 \& 5 respectively. The
fine filamentary nature of the eastern nebulosity is revealed in
Figs. 4 \& 5. At a distance of 5 kpc the finest \oiii\ filaments are
$\approx$~ 7 $\times$ 10$^{16}$cm ($\equiv$ 1\arcsec) wide.

\subsubsection{Deep, high resolution imagery of the northern radio ridge}

The \HNII\ image in Fig. 6 was obtained on the 1 May 2003 on one
position of the extensive northern ridge of radio emission (Dubner et
al. 1998 and shown here in Fig. 1).  For this purpose the Manchester
echelle spectrometer (MES--SPM; \citealt{mea84,mea03}) in its imaging
mode was combined with the 2.1--m San Pedro Martir (Mexico) telescope
(see Sect 2.2.2 for plate scale). The integration time was 2400 s and
a SITe CCD (2$\times$2 binned) was the detector. Faint nebular
filaments parallel to the radio ridge, can be seen for the first
time. No star subtraction was possible in this case for a continuum
image was not obtained. 
  
\subsection{Spectroscopy}  
\subsubsection{Low dispersion - eastern and western filaments}  
  
Low dispersion long--slit spectra were obtained with the 1.3 m
telescope at Skinakas Observatory in June 14, 2004 and September 6--7,
2005. The 1300 line mm$^{-1}$~grating was used in conjunction with a
2000$\times$800 SITe CCD (15$\times$15 $\mu$m$^{2}$~pixels) resulting
in a scale of 1 \AA\ pixel$^{-1}$ and covers the range of 4750 \AA\ --
6815 \AA. The spectral resolution is $\sim$8 pixels and $\sim$11
pixels full width at half maximum (fwhm) for the red and blue
wavelengths, respectively. The slit width is 7.7 arcsec and in all cases
was oriented in the south--north direction; the slit length is
7.9 arcmin. The spectrophotometric standard stars HR5501, HR7596,
HR9087, HR718, and HR7950 \citep{ham94} were observed to calibrate the
spectra.
  
The deep low resolution spectra were taken on the relatively bright
optical filament in the eastern and western parts of W~50 (their exact
positions are given in Table~\ref{table1}). In Table~\ref{sfluxes}, we
present the relative line fluxes taken from three different apertures
(I, II and III) along each slit. In particular, apertures I, II and
III have an offset (see Table~\ref{table1}) north or south of the slit
centre which were selected because they are free of field stars in an
otherwise crowded field and they include sufficient line emission to
permit an accurate determination of the observed line fluxes. The
background extraction aperture was taken towards the northern end or
the southern ends of the slits depending on the slit position.  The
signal to noise ratios presented in Table~\ref{sfluxes}~do not include
calibration errors, which are less than 10 percent. Typical spectra
from the eastern and western filaments are shown in Fig. 7.
\par  
Interstellar reddening was derived from the \ha/\hbeta\ ratio
\citep{ost89}, using the interstellar extinction law by \citet{fit99}  
and R$_{\rm v} = 3.1$. Therefore, the interstellar logarithmic
extinction coefficient c(\hbeta) can be derived by using the
relationship
   
\begin{equation}  
{\rm c(H\beta)} = \frac{1}{0.348} \log\frac{{\rm F(H\alpha)}/{\rm  
F(H\beta)}}{2.85}  
\end{equation}  
   
where, 0.348 is the relative logarithmic extinction coefficient for
\hbeta/\ha\ and 2.85 the theoretical value of F(\ha)/F(\hbeta). Here
we have used the ratio of 2.85 though Hartigan et al. (1987) suggests
3.0 which could be more appropriate when some collisional excitation
is present. The observational reddening in magnitude E$_{\rm
B-V}$~was also calculated using the relationship \citep{sea79}
   
\begin{equation}  
{\rm c(H\beta) = 0.4 \cdot X_{\beta} \cdot E_{\rm B-V}}  
\end{equation}  
   
where the extinction parameter X$_{\beta}=$ 3.615 \citep{fit99}.  
   
The errors on the measurements of c(\hbeta) and E$_{\rm B-V}$~were
calculated through standard error propagation of equations (1) and
(2). Consequently, c was found to be between 0.53 and 2.00 (to give
A$_{\rm V}$~of 1.1 to 4.2) and E$_{\rm B-V}$ between 0.4 and 1.4. 
  
\subsubsection{High dispersion - eastern and western filaments}  
  
Spatially resolved, longslit spectra were obtained of the eastern
filamentary nebulosity on 2--4 August, 2005 with the Manchester
echelle spectrometer (MES--SPM; \citealt{mea84,mea03}) combined with
the 2.1--m San Pedro Martir (Mexico) telescope.  The slit, orientated
EW, was 300 $\mu$m wide ($\equiv$ 3.9 arcsec and 20
\kms). The 512 increments of the 2$\times$2 binned SITe CCD detector, each
0.624 arcsec long, give a total projected slit length of 5.32 arcmin
on the sky. In this spectroscopic mode MES--SPM has no
cross--dispersion consequently, for the present observations, a filter
of 90-\AA\ bandwidth was used to isolate the 87$^{\rm th}$ echelle
order containing the \ha\ and \niiab\ nebular emission lines.
Integration of 1800 s and 3600 s were obtained respectively for slit
positions 1 and 2 marked in Fig. 4. The position--velocity (pv) arrays
of \nii\ line profiles from slit positions 1 \& 2 are shown in Figs. 8
\& 9 respectively. The \nii\ line profiles from the incremental
lengths marked in Figs. 8 \& 9 are shown in Figs. 10 \& 11
respectively.
 
It can be seen in Figs. 7--11 that the brightest filaments (section A
in Figs. 8 \& 10 and section C in Figs. 9 \& 11) emit 22 \kms\ wide
profiles when corrected for instrumental broadening, both centered on
\vhel\ = 56 \kms. This value is significantly different from  
\vsys\ $\approx$ 40 which is the mean of the estimation of 28 \kms\ from the
\hi\ measurements of \citet{dub98} and 52 \kms\ from optical
profiles of the eastern and western filaments (Figs 2 \& 3) given by
Mazeh et al (1983). In the present paper it has been assumed that for
W~50, \vhel\ = \vlsr\ - 14.3 \kms.
 
However, the fainter \nii\ emitting regions between the bright eastern
filaments exhibit approaching radial velocities of up to 100 \kms\
with respect to \vhel\ = 56 \kms. These faint but extensive
high--speed regions are particularly prominent for the sections D and
B in Fig. 9 whose profiles are shown in Fig. 11.

These high dispersion spectral observations of the eastern filamentary
nebulosity were obtained in non-photometric conditions and were not
therefore compared photometrically with a standard star.

\subsubsection{High dispersion -- northern ridge}

Nearly the same instrumental setup and data analysis as described in
Sect. 2.2.2 was employed on 1 May 2003 for the longslit spectroscopy
of the northern nebulosity shown in Fig. 6. Here the slit was now 150
$\mu$m wide ($\equiv$ 1.9 arcsec and 9 \kms), orientated NS and
centred on RA 19h 11m 00.72s, DEC 05$\degree$ 19$\arcmin$ 16$\arcsec$
(J2000). The integration time was 4200 s. The position velocity (pv)
array of \nii\ profiles obtained in this way are shown in Fig. 12 and
the line profile for the section marked A in Fig. 12 is shown in
Fig. 13.

This spectral observation was calibrated photometrically against the
slitless spectrum of the standard star Feige 56 to give a value,
uncorrected for interstellar extinction, for the total emission in
this \nii\ profile of 1.9$\times$10$^{-17}$ erg s$^{-1}$ cm$^{-2}$
arcsec$^{-2}$ to 20 percent accuracy. This is $\approx$ 29 times
fainter in the \nii\ line than the brightest filament in the eastern
nebulosity as listed in Table 1.  The \nii\ profile is centred on
\vhel\ = 56 $\pm$ 2 \kms\ and 36 \kms\ wide when simulated by a single
Gaussian. When corrected for instrumental broadening this width
reduces to 35 \kms.
 
\section{Discussion}

\subsection{Location of optical emission} 
The \ha/\hbeta\ ratios listed in Table 2 strongly indicates that heavy
and patchy absorption of optical emission occurs over W~50. This is to
be expected for a distance of 5 kpc along the Galactic plane and
for a Galactic latitude of only -2.27\degr. This possibility is
further supported by the ISOCAM infrared emission map of the vicinity
of the western filaments obtained by \citet{mol05}. One patch of
infrared emission coincides closely with the visible western filaments
(Fig. 3a \& b and Fig. 5) but a larger region of infrared emission has
no optical counterpart in our images. It is therefore probable that
only parts of the shock--excited optical emission from W~50 is being
observed in Figs. 2--5 with considerable parts heavily obscured by the
foreground dust.

This patchiness by foreground dust as the cause of the limited
detection of the whole of the W~50 optical filaments, which would be
expected if this is indeed a supernova remnant, could be confirmed by
the optical detection in Figs. 6, 12 \& 13 for the first time of the
northern radio ridge. Although the latter's optical emission is very
much fainter than that observed from the eastern and western optical
filaments the \nii\ profile from it in Fig. 13 is centred on \vhel\ =
56 \kms. This matches closely the values in Figs. 10 \& 11 for the
brightest and narrowest profiles of the eastern filamentary
nebulosity. It is suggested that in all of these regions the motions
of the emitting gas are nearly perpendicular to the sight--line and
that \vhel\ = 56 \kms\ could be the best value of the systemic
heliocentric radial velocity, \vsys\ for the whole SS~433 and W~50
complex. In these circumstances the optical emission from the northern
radio ridge would have to be heavily obscured by the patchy dust.
One complication to this patchy dust interpretation 
is that the relativistic jet of SS~433 has injected
$\approx$ 2 $\times$ 10$^{51}$ erg into the surrounding medium over its
lifetime and could have affected the recombinations rates
of the shocked W~50 gas in inpredictable ways.
  
\subsection{Morphology of the eastern and western optical filaments}  
Taken as a whole, the new optical imagery in Figs. 2--5 reveals that
the eastern filaments trace a broad arc that is convex with respect to
the central source and follows the large scale helical morphology of
the W~50 boundary in this vicinity as revealed by radio continuum
maps (\citealt{dub98}; see their more detailed maps to appreciate this
structure). Interestingly, the western filaments seem to curve in an
identical sense despite being on the opposite side of SS~433 (at all
wavelengths the morphology of the western `ear' is most likely
distorted by a strong interaction with a dense portion of the local
ISM).
  
The bi--polar relativistic jet from SS~433 appears to have broken
through diametrically opposite boundaries of the W~50 supernova
remnant assuming that this is the origin of large circular region of
radio emission in Fig. 1. The bright optical emission (Figs. 2 -- 5)
then lies at these breakout regions. In an idealised case, a breakout
region should take the form of a ring of emission where the expanding
jet envelope shocks the dense shell of swept up and compressed
material at the supernova remnant boundary. Optical emission from this
interaction is not expected elsewhere because the interior is too hot
and rarefied and the ambient medium is too low density ahead of the
jet. The observed arc morphology i.e. eastern and western arcs, could
be simply the result of heavy, patchy foreground interstellar
absorption (see Sect. 3.1) combined with the jet axis being tilted so
that the eastern tip is pointing towards the observer and western tip
away (see references in \citealt{fab04} that confirm this jet
orientation); the rings of optical emission become partial ellipses
with only localised portions on their nearsides with respect to the
observer being visible. However, the interaction is likely to be much
more complex than described above not least because the jet close to
SS~433 has a `corkscrew' structure.  Also \citet{vel00} have
numerically modelled the W~50/SS~433 system which in their simulations
results in reflected and secondary shocks as well as the filling and
acceleration of the whole supernova remnant by the jet
cocoon. Nonetheless the breakout region remains a distinct zone in the
simulations.

\subsection{The shocked emission}
All of the spectra in Table~2 show clearly that the observed optical
emission originates in shock heated gas, since the \sii/\ha\ $>$
1.5. The \oiii\ emission detected in both spectra suggests a shock
velocity greater than 100 \vel\ (Cox \& Raymond 1985).  The absolute
\ha\ flux covers a range of values from 1.4 to 2.6 $\times$ \flux\ and
5.8 to 16.3 $\times$ \flux\ for the western and eastern nebulosities,
respectively. The \siirat\ ratio which was found to be between 1.1 and
1.4. 
Using `temden' in the nebular package in IRAF \citep{sha95} the
electron densities of 50 cm$^{-3}$ and 700 cm$^{-3}$ are measured for
the eastern and western filaments, respectively.  A comparison to
shock models shows that the densities are different (east to west) by
a factor of 10 or more, while the shock velocities are nearly the
same.  Hence, either there is a huge pressure differential between the
two sides (nT$\sim$pressure) or there is something incorrect with this
interpretation. Furthermore, \hbeta\ emission was detected in both
areas (but with low S/N) to give \oiii/\hbeta\ ratios of between 5.6
(area EI) and 15.7 (area WII).  Theoretical models of \citet{cox85}
and \citet{har87} suggest that for shocks with complete recombination
zones the expected \oiii/\hbeta\ ratio is $\sim$6, while this limit is
exceeded in the case of a shock with incomplete recombination zones
\citep{ray88}. Our measured values suggest in the western nebulosity
that shocks with incomplete recombination zones are present while in
the eastern nebulosity the presence of shocks with complete
recombination zones could not be ruled out.  A combination therefore
of the present observations with the theoretical predictions suggests
that shock velocities are $\simeq$100 \vel\ and $\simeq$120 \vel\ for
the eastern and western nebulosities respectively, though this
difference may not be real when uncertainties in the data are
considered.
  
\subsection{Velocity of the optical filaments}  
The previous optical spectra of \citet{maz83} at very low angular
resolution of the eastern and western filaments suggest that the
average heliocentric radial velocities (\vhel) are, respectively, 65
and 40~\kms\ (where the LSR and heliocentric radial velocities are
related by \vhel\ $=$ \vlsr\ -14.3 \kms). This is in apparent
contradiction with the orientation of the jet, which has the eastern
portion directed towards the observer. The radio morphology also
indicates that the axis of the elongated lobes of the W 50 shell is
tilted to the plane of the sky with the eastern side nearest the
observer.

The present spectral observations, with their higher spatial
resolution, show that the brightes regions of the eastern filaments
(Figs. 8--11) emit the narrowest lines centred on \vhel\ = 56 \kms\
but with extensive fainter regions, flowing off the filaments
(Fig. 9), composed of high--velocity gas approaching the observer with
radial velocities continuously up to 100 \kms\ from the bright
filament value. It remains possible that similar high spatial
resolution observations of the western filaments will show similar
complex motions and eliminate this discrepancy which could only be a
consequence of the low angular resolution employed in the early
measurements. However, the appropriate flows emitting faintly in the
eastern filaments are consistent with eastern side of W~50 pointing
towards the observer.
  
Also, as noted above, the expected local shock velocities that give
rise to the optical filaments are of order 100~\kms\ and the expansion
of the remnant as a whole is estimated by Dubner et al (1998) to be
$\sim$~75~\kms. Thus the measured difference in radial velocity
between the bright optical filaments is small compared with the
expected velocity range ($\sim$ 100~\kms). Any residual difference
could be explained easily by patchy dust absorption hiding the full
extent of the optical emission.
  
Also the contradiction only exists if it is assumed that the optical
emission is tracing directly the outflow from the star. However, the
optical emission much more likely traces the interaction between the
jet cocoon and the shell of the W~50 SNR. Expansion of this shocked
optically emitting gas could result in localised flows towards and/or
away from the star (cf. simulations of \citealt{vel00}, their figure
3). In the case of the eastern filaments, the curved morphology could
then be a result of an expansion back towards SS~433 whereas the
western filaments could be a result of an expansion away from or
stationary with respect to SS~433. The optical emission then, is
concentrated at the ``rims" of the breakout region.

Broader, competing, possibilities should also be considered for it has
been suggested as above that these elongated breakout features in the
radio map are the microquasar jets currently penetrating through the
shell of the W~50 SNR (Dubner et al., 1998). Alternatively, these
eastern and western radio lobes of the SNR may simply be revealing the
imprint of the precessing jets of SS~433, created early in the
object's evolution after the explosion of the supernova, and which
have inflated along with the rest of the SNR as it expanded. The faint
$\geq$~100~\kms\ outflows from the eastern filaments around this
lobe's apparent breakout region would favour the first possibility.
  
\section*{Acknowledgements}  
  
The authors would like to thank the referee for contructive comments
that have improved the paper considerably. We also thank G. Dubner who
kindly provided us the radio image of W~50 in FITS format and the
staff at Skinakas and SPM Observatories for their excellent support
during these observations. JA and SA acknowledge funding by the
European Union and the Greek Ministry of Development in the framework
of the programme `Promotion of Excellence in Research Institutes (2nd
Part)'. JAL acknowledges financial support from UNAM grants IN 112103,
108406 and 108506. Skinakas Observatory is a collaborative project of
the University of Crete, the Foundation for Research and
Technology-Hellas and the Max-Planck-Institut f\"ur Extraterrestrische
Physik.

%
 
  
\begin{table*}  
\caption[]{Imaging and Spectral log}  
\label{table1}
\begin{flushleft} 
\begin{tabular}{lcccccc}  
\noalign{\smallskip}  
\hline  
\multicolumn{6}{c}{IMAGING} \\  
\hline
Filter &
$\lambda_{\rm c}$ &
$\Delta \lambda$ & Exp. time &  Area & Observatory \\
 & ($\AA$) & ($\AA$) & (sec) & & \\
\hline
\HNII & 6570 & 75 & 2400  & W50 total (4)$^{\rm a}$ & 0.3m Skinakas \\
\oiii & 5005 & 28 & 2400 & W50 total (4) & 0.3m Skinakas \\
Cont blue & 5470 & 230 & 180 & W50 total (4) & 0.3m Skinakas \\
Cont red & 6096 & 134 & 180 & W50 total (4) & 0.3m Skinakas \\
\oiii & 5005 & 28 & 2400 & East (10) & 1.3m Skinakas \\
Cont blue & 5470 & 230 & 180 & East (10) & 1.3m Skinakas \\
\HNII & 6570 & 75 & 2400 & West (2) & 1.3m Skinakas \\
Cont red & 6096 & 134 & 180 & West (2) & 1.3m Skinakas \\
\HNII & 6580 & 90 & 2400 & North (1) & 2.1m SPM \\
\hline
\multicolumn{7}{c}{SPECTROSCOPY} \\  
\hline  
Area & \multicolumn{2}{c}{Slit centres} & Exp. time & Offset$^{\rm b}$ & Aperture length$^{\rm c}$ & Observatory \\  
 & $\alpha$ & $\delta$ & (sec) & (arcsec) & (arcsec)& \\  
\hline  
East I (EI) & 19\h14\m20\s & 05\degr03\arcmin41\arcsec  & 3900 & 112 S & 12 & 1.3m Skinakas\\  
East II (EII) & 19\h14\m20\s & 05\degr03\arcmin41\arcsec  & 3900 &  94 S & 26 & 1.3m Skinakas\\   
East III (EIII) & 19\h14\m20\s & 05\degr03\arcmin41\arcsec  & 3900 & 61 S & 66 & 1.3m Skinakas\\  
East IV (EIV) & 19\h14\m24\s & 05\degr04\arcmin49\arcsec & 3900 & 58 S & 30 & 1.3m Skinakas\\  
East V (EV) & 19\h14\m38\s & 04\degr45\arcmin40\arcsec & 3900 & 4 N & 24 & 1.3m Skinakas\\  
West I (WI) & 19\h09\m39\s & 05\degr02\arcmin34\arcsec  & 3900 & 89 S & 36 & 1.3m Skinakas\\  
West II (WII) & 19\h09\m39\s & 05\degr02\arcmin34\arcsec  & 3900 & 50 S & 43 & 1.3m Skinakas\\  
West III (WIII) & 19\h09\m39\s & 05\degr02\arcmin34\arcsec & 3900 & 193 N & 44  & 1.3m Skinakas\\
East slit 1 & 19\h14\m18\s & 05\degr02\arcmin00\arcsec & 1800 & 0$^{\rm d}$ & 300$^{\rm e}$ & 2.1m SPM\\
East slit 2 & 19\h13\m57\s & 04\degr55\arcmin30\arcsec & 3600 & 0$^{\rm d}$ & 300$^{\rm e}$ & 2.1m SPM\\
North slit & 19\h11\m01\s & 05\degr19\arcmin16\arcsec & 4200 & 90$^{\rm d}$ & 150$^{\rm e}$ & 2.1m SPM\\ 
\hline  
\end{tabular}
\end{flushleft}
\begin{flushleft}
${\rm ^a}$ Numbers in parentheses represent the total number of different fields.\\ 
${\rm ^b}$ Spatial offset from the slit centre in arcsec: N($=$North), S($=$South).\\  
${\rm ^c}$ Aperture lengths for each area in arcsec.\\ 
${\rm ^d}$ Slit position angle (PA) in degrees.\\ 
${\rm ^e}$ Slit width in $\mu$m.\\ 
\end{flushleft} 
\end{table*}  
\begin{table*}  
\caption[]{Relative line fluxes.}  
\label{sfluxes}
\begin{flushleft}  
\begin{tabular}{llllllllllllllll}  
\hline  
 \noalign{\smallskip}  
 & \multicolumn{15}{c}{East} \\  
 & \multicolumn{3}{c}{Area EI} & \multicolumn{3}{c}{Area EII} & \multicolumn{3}{c}{Area EIII} & \multicolumn{3}{c}{Area EIV} & \multicolumn{3}{c}{Area EV}\\  Line (\AA) & F$^{\rm a}$ &  I$^{\rm b}$ &  S/N$^{\rm c}$ & F &  I &  S/N & F &  I &  S/N & F &  I &  S/N & F &  I &  S/N \\  
\hline  
\hbeta\ 4861 & 16 & 35 & (7) & 11 & 35 & (6) & 15 & 35 & (5) & 7 & 35 & (1) & 23 & 35 & (4) \\  
\oxygen\ 4959 & 25 & 52 & (9) & 23 & 68 & (13) & 22 & 49 & (8) & $-$ & $-$ & $-$ & $-$ & $-$ & $-$  \\  
\oxygen\ 5007 & 66 & 134 & (27) & 61 & 174 & (31) & 84 & 181 & (25) & 157 & 488 & (28) & 70 & 103 & (14)  \\  
\oi\ 6300 & 67 & 74 & (38) & 78 & 90 & (61) & 87 & 97 & (38) & 34 & 40 & (13) & 44 & 46 & (19) \\  
\oi\ 6363 & 17 & 18 & (15) & 22 & 25 & (22) & 26 & 28 & (16) & $-$ & $-$ & $-$ & 45 & 47 & (20)\\  
\nitrogen\ 6548 & 106 & 107 & (70) & 116 & 117 & (90) & 105 & 106 & (61) & 75 & 76 & (29) & 101 & 102 & (43) \\  
\ha\ 6563 & 100 & 100 & (65) & 100 & 100 & (81) & 100 & 100 & (55) & 100 & 100 & (38) & 100 & 100 & (41) \\  
\nitrogen\ 6584 & 343 & 340 & (154)& 377 & 373 & (210)& 370 & 367 & (159)& 228 & 225 & (78) & 316 & 315 & (109) \\  
\sulfur\ 6716 & 128 & 121 & (80) & 139 & 128 & (110)& 172 & 162 & (88) & 50 & 46 & (24) & 140 & 136 & (56)\\  
\sulfur\ 6731 & 89 & 84 & (61) & 98 & 90 & (81) & 122 & 114 & (63) & 33 & 30 & (15) & 102 & 99 & (41) \\
\hline
Absolute \ha\ flux$^{\rm d}$ & \multicolumn{3}{c}{16.3} & \multicolumn{3}{c}{14.7} & \multicolumn{3}{c}{5.8} & \multicolumn{3}{c}{2.9} & \multicolumn{3}{c}{4.9} \\  
\sulfur/\ha\ & \multicolumn{3}{c}{2.17$\pm$0.06} & \multicolumn{3}{c}{2.36$\pm$0.05} & \multicolumn{3}{c}{2.93$\pm$0.09} & \multicolumn{3}{c}{0.83$\pm$0.06} & \multicolumn{3}{c}{2.42$\pm$0.11} \\  
F(6716)/F(6731) & \multicolumn{3}{c}{1.43$\pm$0.11} & \multicolumn{3}{c}{1.42$\pm$0.03} & \multicolumn{3}{c}{1.41$\pm$0.04} & \multicolumn{3}{c}{1.49$\pm$0.18} & \multicolumn{3}{c}{1.38$\pm$0.06} \\  
\oxygen/\hbeta\  & \multicolumn{3}{c}{5.27$\pm$0.49} & \multicolumn{3}{c}{6.83$\pm$0.46} & \multicolumn{3}{c}{6.50$\pm$0.73} & \multicolumn{3}{c}{13.74$\pm$4.03} & \multicolumn{3}{c}{2.91$\pm$0.62} \\   
c(\hbeta) & \multicolumn{3}{c}{0.98$\pm$0.12} & \multicolumn{3}{c}{1.45$\pm$0.13} & \multicolumn{3}{c}{1.06$\pm$0.15} & \multicolumn{3}{c}{1.57$\pm$0.37} & \multicolumn{3}{c}{0.53$\pm$0.17} \\ 
E$_{\rm B-V}$ & \multicolumn{3}{c}{0.68$\pm$0.08} & \multicolumn{3}{c}{1.00$\pm$0.09} & \multicolumn{3}{c}{0.73$\pm$0.10} & \multicolumn{3}{c}{1.09$\pm$0.26} & \multicolumn{3}{c}{0.37$\pm$0.12} \\    
\hline
 & \multicolumn{9}{c}{West} \\  
 & \multicolumn{3}{c}{Area WI} & \multicolumn{3}{c}{Area WII} & \multicolumn{3}{c}{Area WIII}\\
Line (\AA) & F &  I &  S/N & F &  I &  S/N & F &  I &  S/N\\  
\hline  
\hbeta\ 4861 & 14 & 35 & (2) & 7 & 35 & (2) & 14 & 35 & (3) \\  
\oxygen\ 4959 & 28 & 66 & (2) & 23 & 104 & (3) & 37 & 88 & (4) \\  
\oxygen\ 5007 & 89 & 204 & (6) & 87 & 373 & (9) & 159 & 364 & (10) \\  
\nitrogen\ 6548 & 79 & 80 & (10) & 87 & 88 & (18) & 75 & 76 & (10) \\  
\ha\ 6563 & 100 & 100 & (12) & 100 & 100 & (20) & 100 & 100 & (12) \\  
\nitrogen\ 6584 & 257 & 255 & (29) & 275 & 271 & (52) & 301 & 298 & (33) \\  
\sulfur\ 6716 & 82 & 77 & (11) & 120 & 107 & (25) & 112 & 105 & (14) \\  
\sulfur\ 6731 & 74 & 69 & (10) & 92 & 81 & (20) & 107 & 100 & (14) \\    
\hline
Absolute \ha\ flux & \multicolumn{3}{c}{1.4} & \multicolumn{3}{c}{2.6} & \multicolumn{3}{c}{1.4} \\  
\sulfur/\ha\ &  \multicolumn{3}{c}{1.56$\pm$0.27} & \multicolumn{3}{c}{1.86$\pm$0.19} &  \multicolumn{3}{c}{2.12$\pm$0.33} \\  
F(6716)/F(6731) & \multicolumn{3}{c}{1.11 $\pm$0.23} & \multicolumn{3}{c}{1.31$\pm$0.14} &  \multicolumn{3}{c}{1.21$\pm$0.16} \\  
\oxygen/\hbeta\  &  \multicolumn{3}{c}{7.64$\pm$2.32} & \multicolumn{3}{c}{13.40$\pm$1.81} & \multicolumn{3}{c}{12.78$\pm$2.40} \\   
c(\hbeta) &  \multicolumn{3}{c}{1.15$\pm$0.67} & \multicolumn{3}{c}{2.01$\pm$0.76} & \multicolumn{3}{c}{1.15$\pm$0.65} \\ 
E$_{\rm B-V}$ & \multicolumn{3}{c}{0.80$\pm$0.46} & \multicolumn{3}{c}{1.39$\pm$0.52} & \multicolumn{3}{c}{0.80$\pm$0.45} \\ 
\hline  
\end{tabular}
\end{flushleft}  
\begin{flushleft}   
   
${\rm ^a}$ Observed fluxes normalised to F(H$\alpha$)=100 and uncorrected  
for interstellar extinction.  

${\rm ^b}$ Intrinsic fluxes normalised to F(H$\alpha$)=100 and corrected  
for interstellar extinction. 

${\rm ^c}$ Numbers in parentheses represent the signal to noise ratio of the
quoted fluxes.\\ 

$^{\rm d}$ In units of \flux.  

Listed fluxes are a signal to noise weighted average of two fluxes.\\  
The emission line ratios \sulfur/\ha, F(6716)/F(6731) and \oxygen/\hbeta\ are calculated using the corrected for interstellar extinction values.\\

The errors of the emission line ratios, c(\hbeta) and E$_{\rm B-V}$ are calculated through standard error propagation.\\
\end{flushleft}  
\end{table*}

 
\newpage  
  
\begin{figure*}  
\centering  
\scalebox{0.90}{\includegraphics{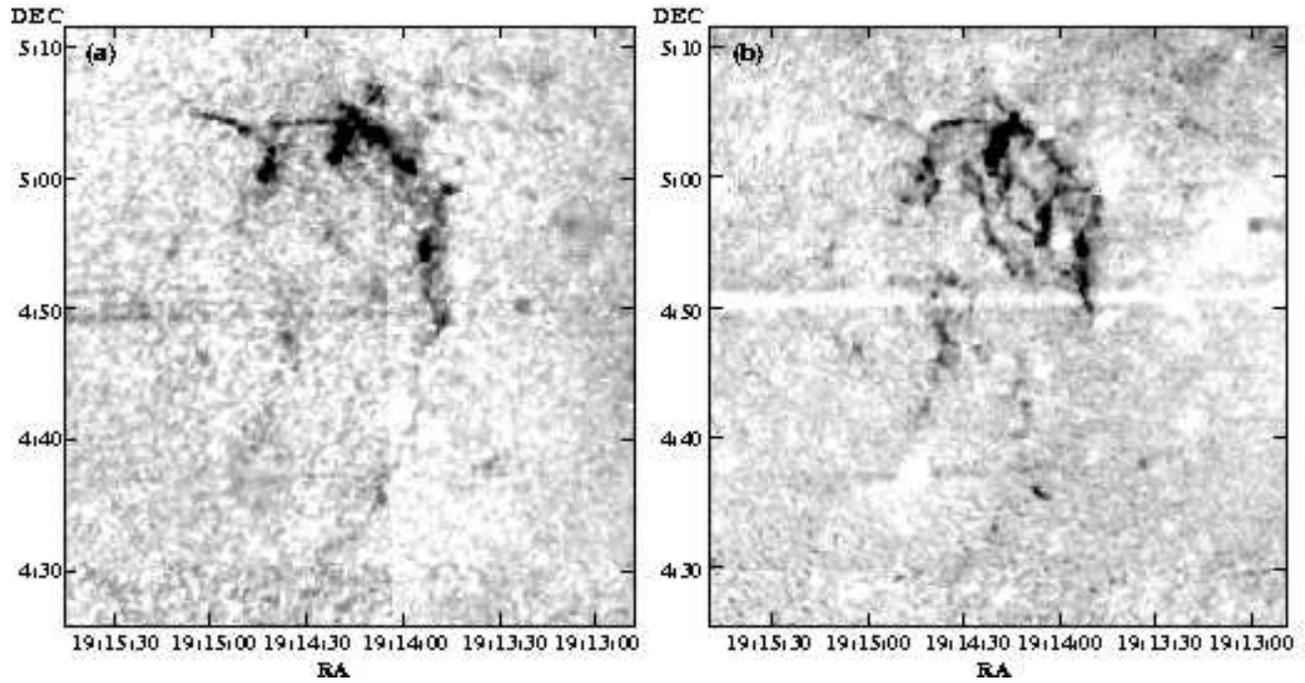}}  
\caption[]{The eastern complex filamentary structure in the light of (a)   
\oiii\ and (b) \hnii. The image has been smoothed to
suppress the residual from the imperfect continuum subtraction.}  
\label{fig02}  
\end{figure*}  
  
\begin{figure*}  
\centering  
\scalebox{0.90}{\includegraphics{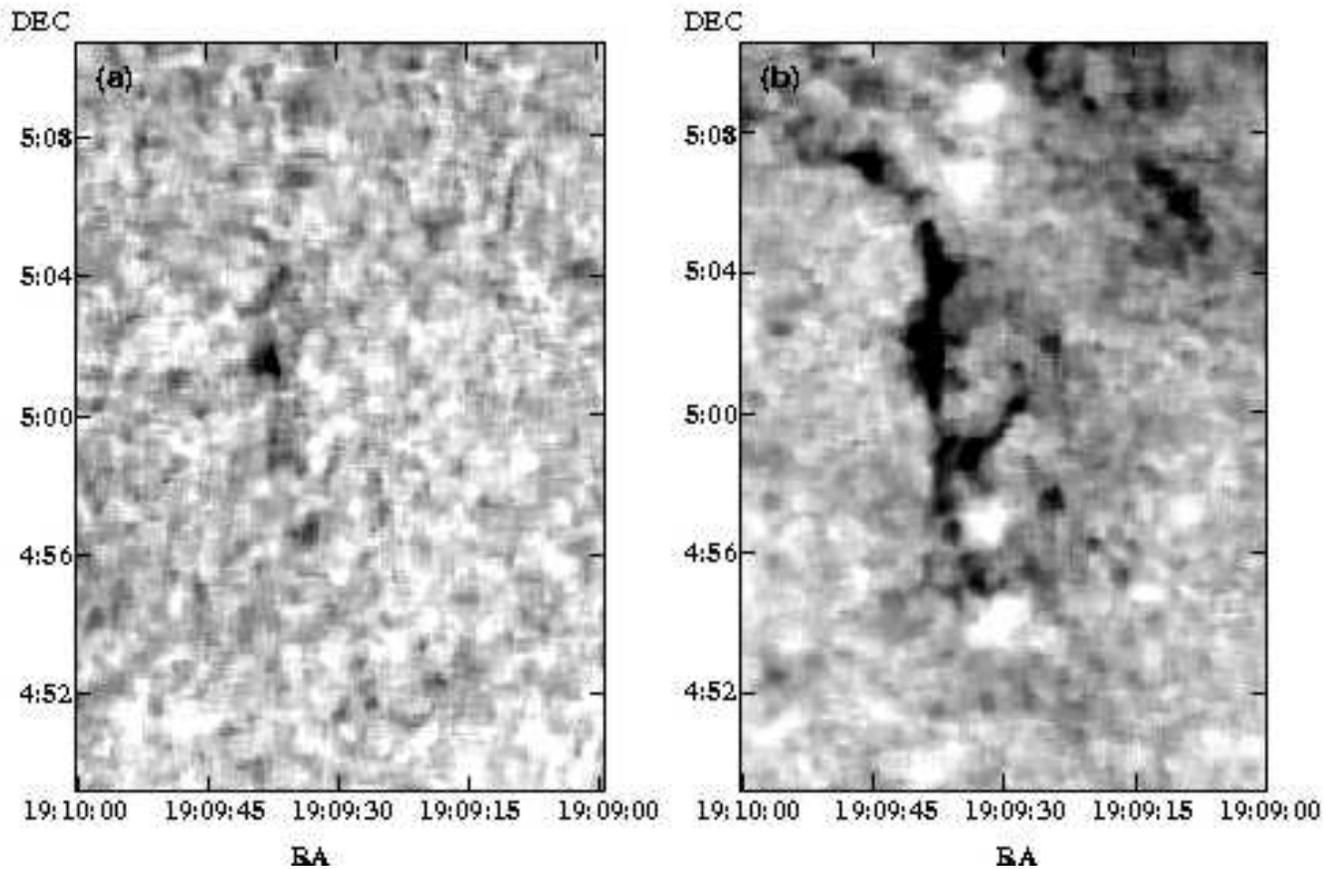}}  
\caption[]{The western filaments in the light of (a) \oiii\ and (b)
\hnii. The image has been smoothed to suppress the residual from the
imperfect continuum subtraction.}
\label{fig03}  
\end{figure*}  
  
\begin{figure*}  
\centering  
\scalebox{1.20}{\includegraphics{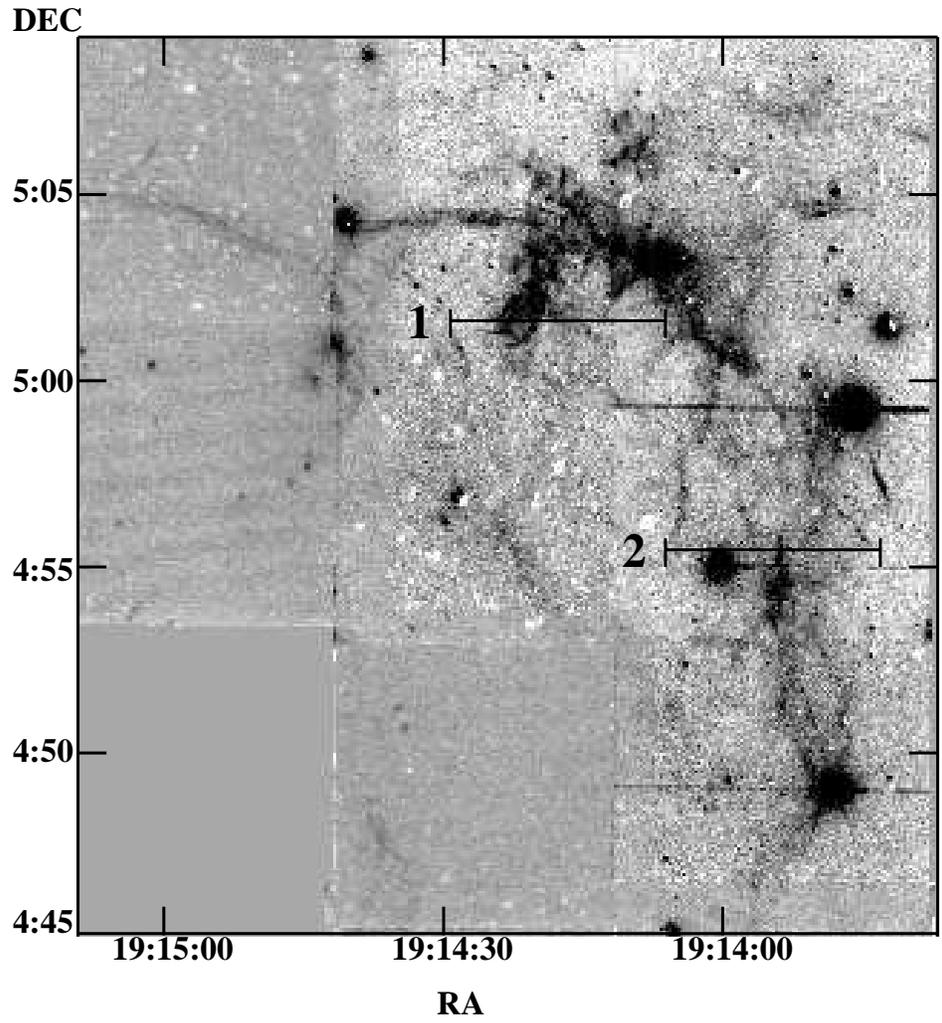}}  
\caption[]{The eastern complex taken with the 1.3m telescope in the
light of \oiii. The positions of slits 1 \& 2 are marked. The image
has been smoothed to suppress the residual from the imperfect
continuum subtraction.}
\label{fig04}  
\end{figure*}  

\begin{figure*}  
\centering  
\scalebox{1.20}{\includegraphics{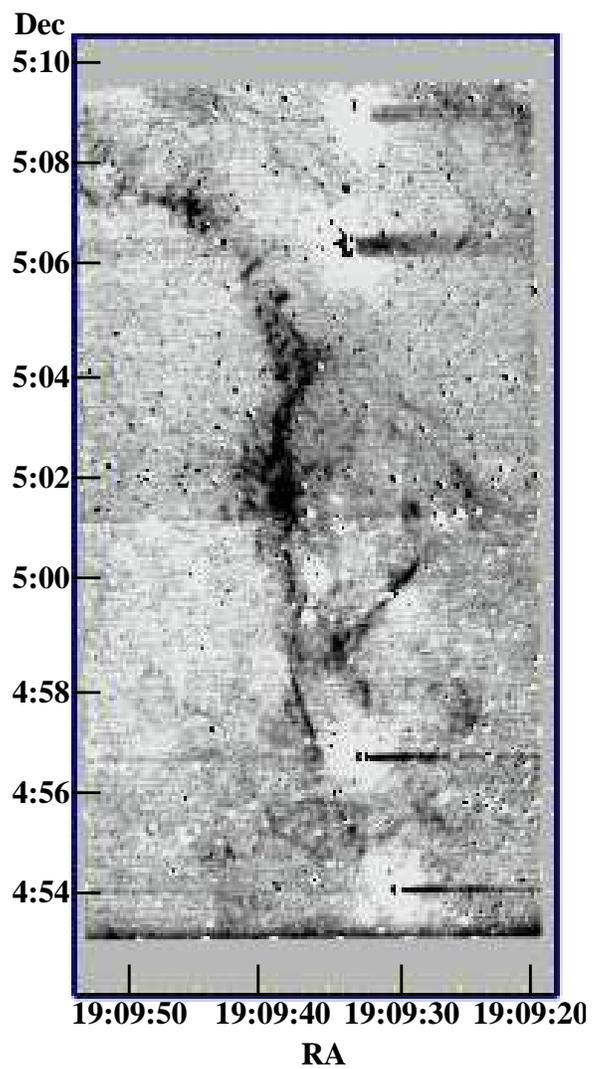}}  
\caption[]{The western complex taken with the 1.3m telescope in the
light of \HNII. The image has been smoothed to suppress the residual
from the imperfect continuum subtraction.}
\label{fig05}  
\end{figure*}  

\begin{figure*}  
\centering  
\scalebox{1.00}{\includegraphics{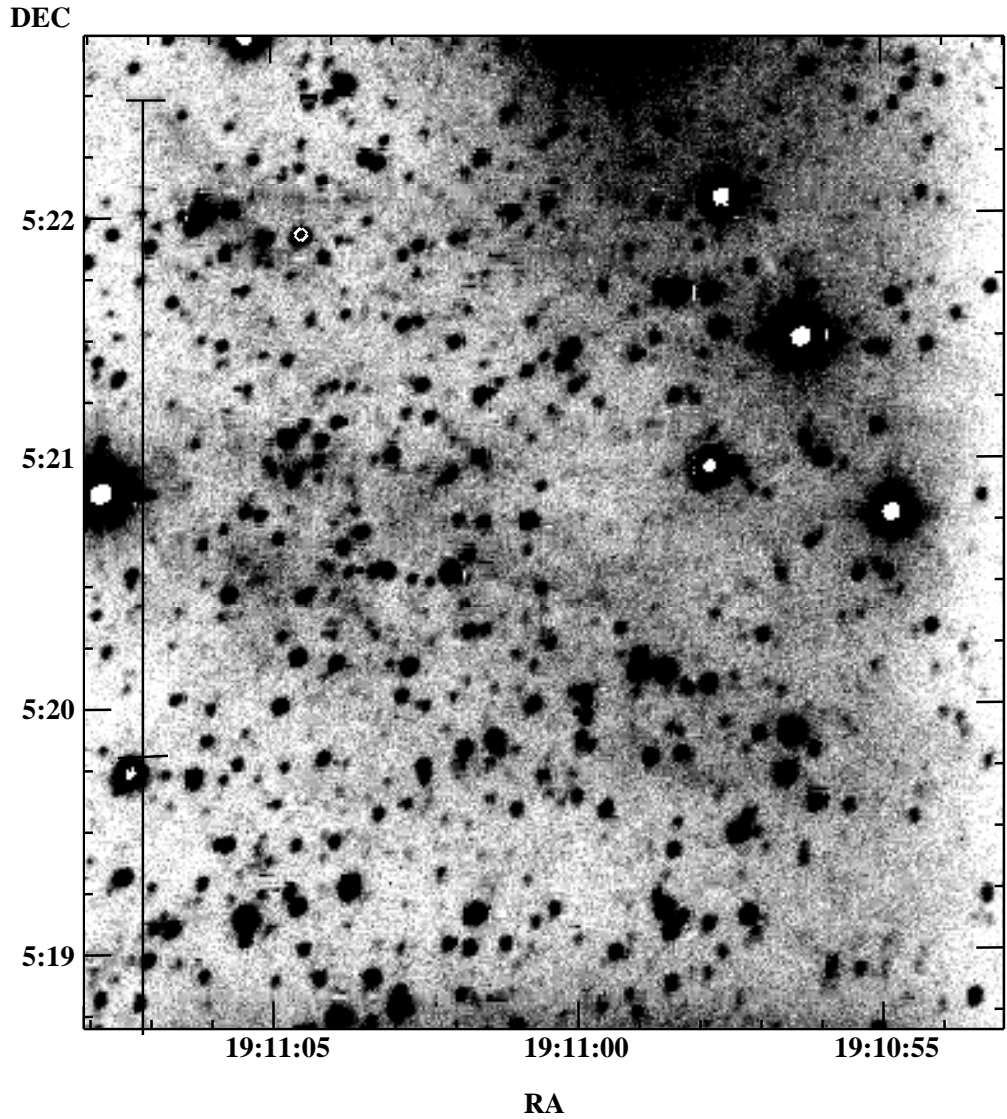}}  
\caption[]{The northern ridge in the light of \hnii. The slit position
for the position velocity array shown in Fig. 12 is marked as a vertical
line. The slit centre is indicated by a horizontal line for the slit
is longer than this image. The image was not continuum subtracted and
attempts to remove star images by the Starlink routine PATCH left
too many confusing residuals so was not pursued. The dark patch at the
top of the image is caused by the halo of a bright star image just off the
field.}
\label{fig06}  
\end{figure*}  
   
\begin{figure*}  
\centering  
\scalebox{0.70}{\includegraphics{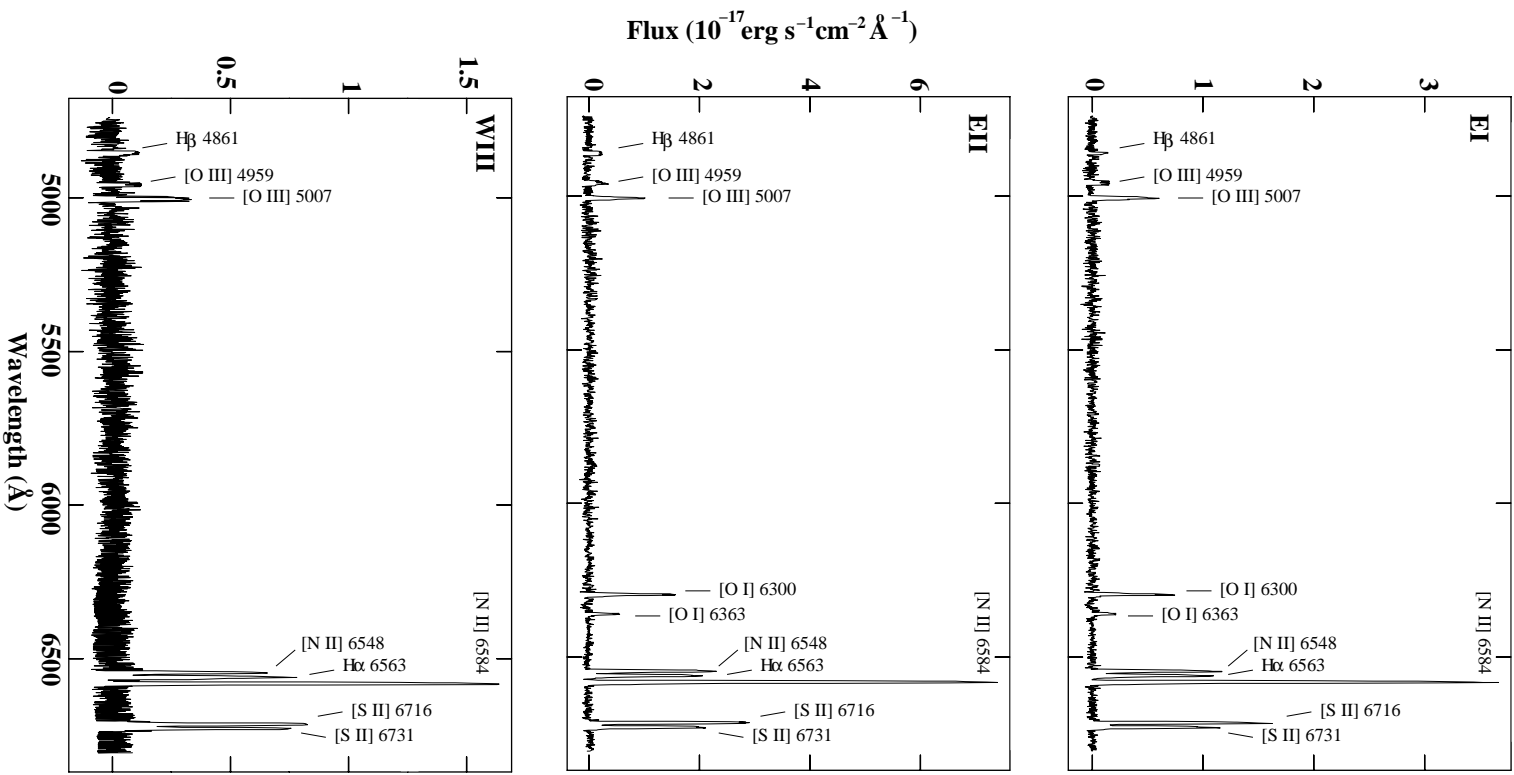}}  
\caption[]{Typical observed spectra from different positions of W50   
(see Table~\ref{sfluxes}).}  
\label{fig07}  
\end{figure*}  
  
\begin{figure*} 
\centering 
\scalebox{0.90}{\includegraphics{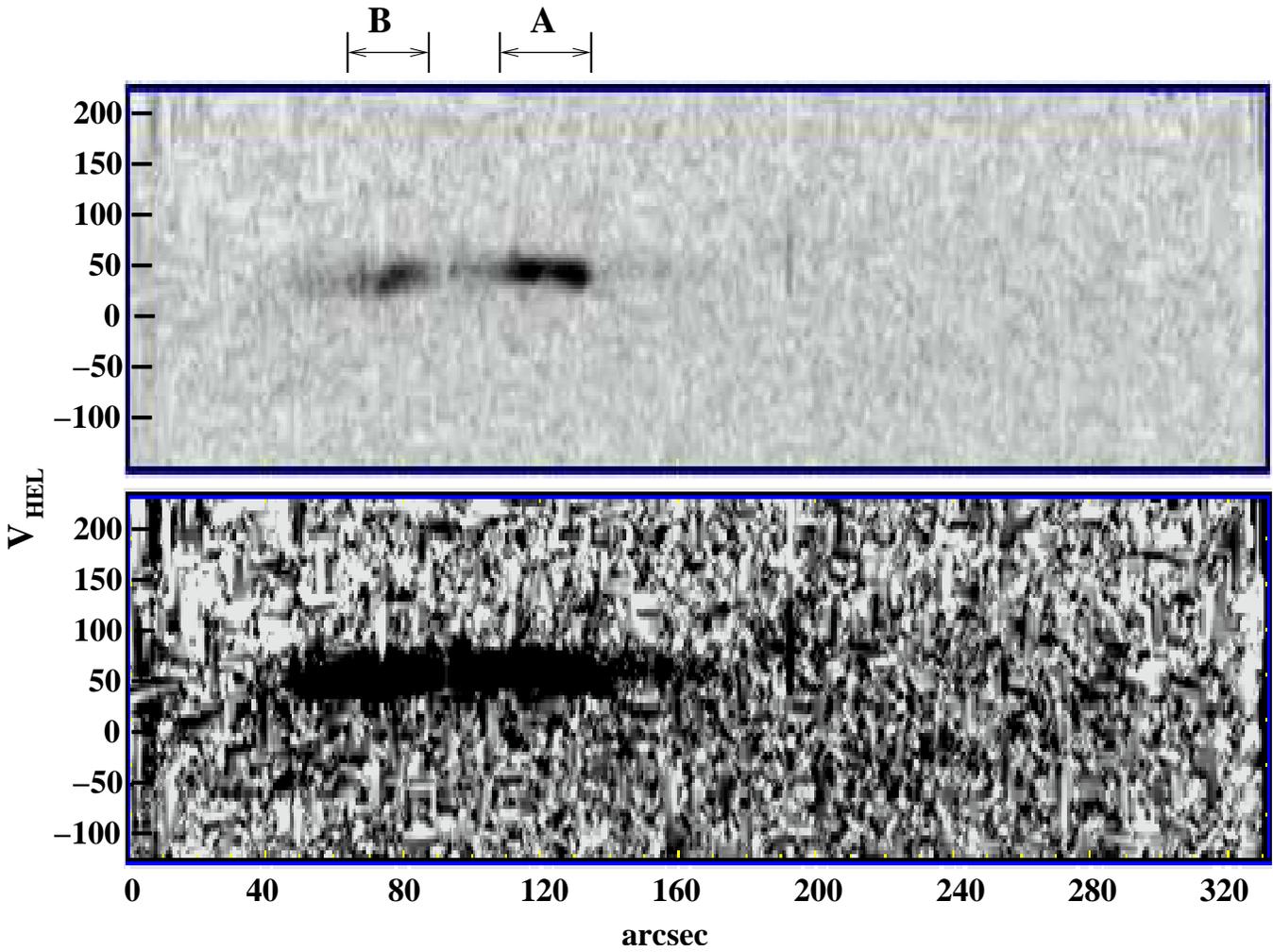}} 
\caption[]{Light and dark representations of the position velocity
array of \nii\ line profile from slit position 1 (see Fig. 4).}
\label{fig8} 
\end{figure*} 
 
\begin{figure*} 
\centering 
\scalebox{0.90}{\includegraphics{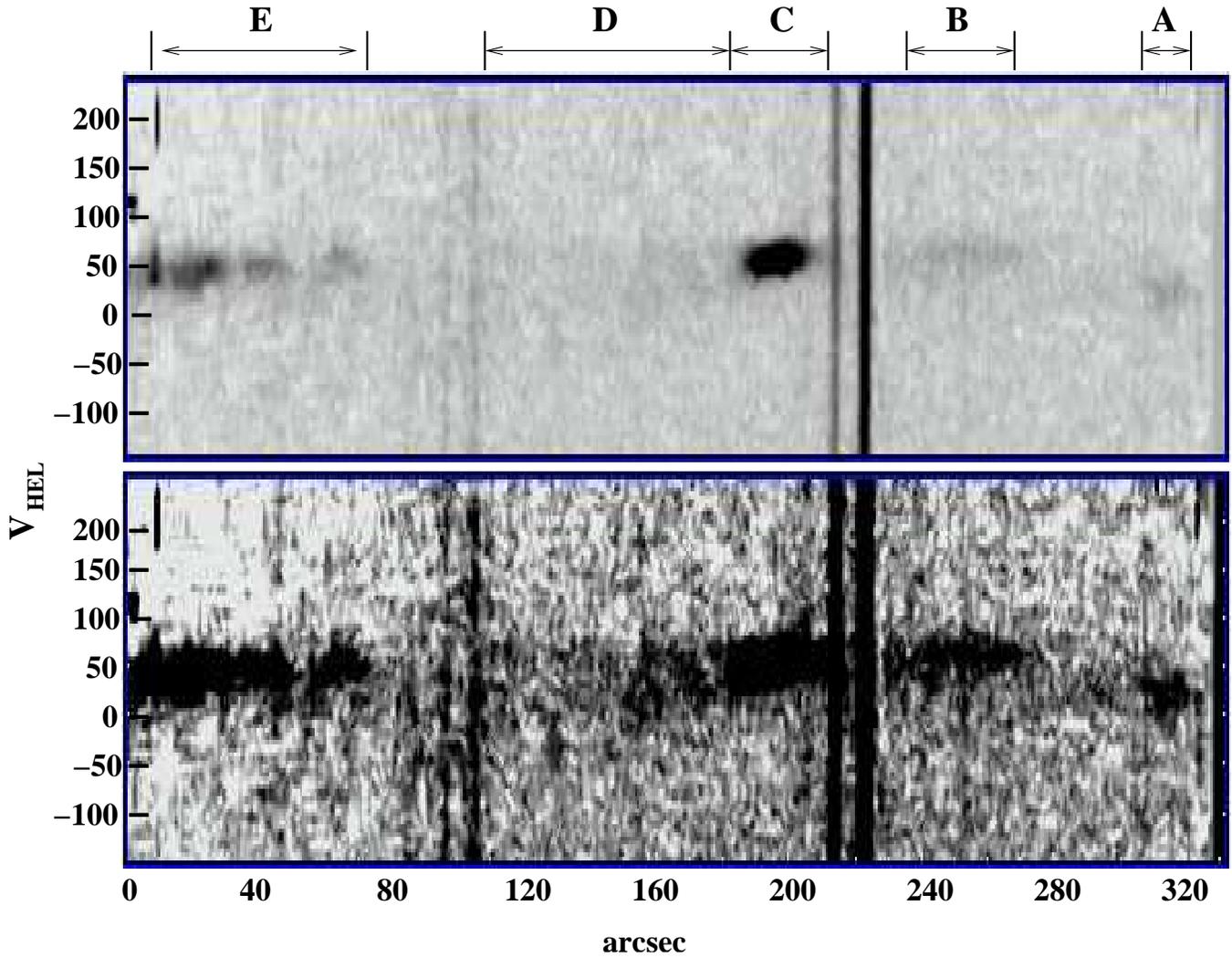}} 
\caption[]{As for Fig. 8 but for slit position 2. The vertical
dark lines are the continuous spectra of faint field stars also
intercepted by the slit.} 
\label{fig9} 
\end{figure*} 
 
\begin{figure*} 
\centering 
\scalebox{1.0}{\includegraphics{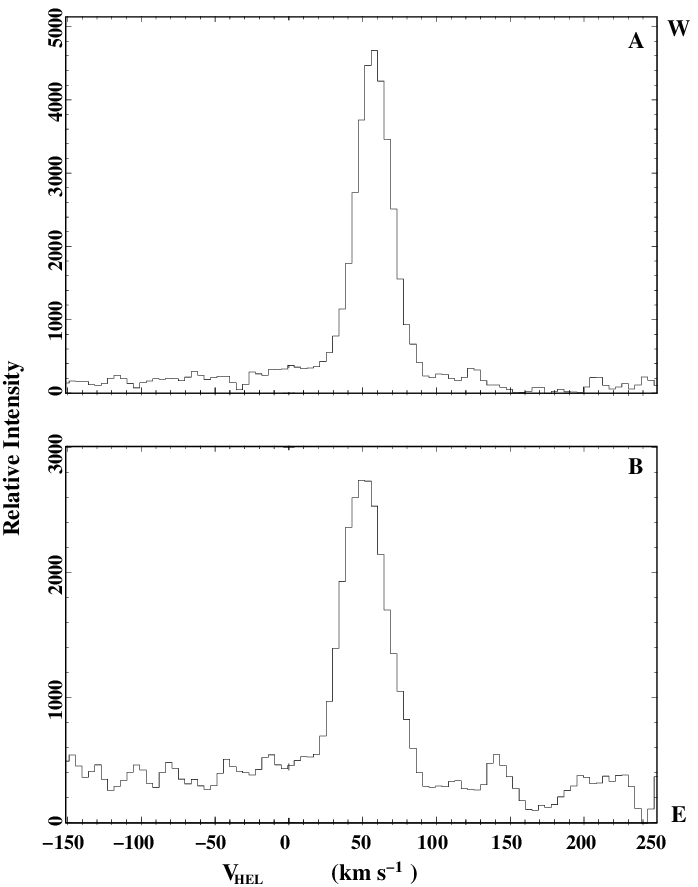}} 
\caption[]{The \nii\ line profiles from the incremental lengths marked
in Fig. 8.}
\label{fig10} 
\end{figure*} 
 
\begin{figure*} 
\centering 
\scalebox{1.0}{\includegraphics{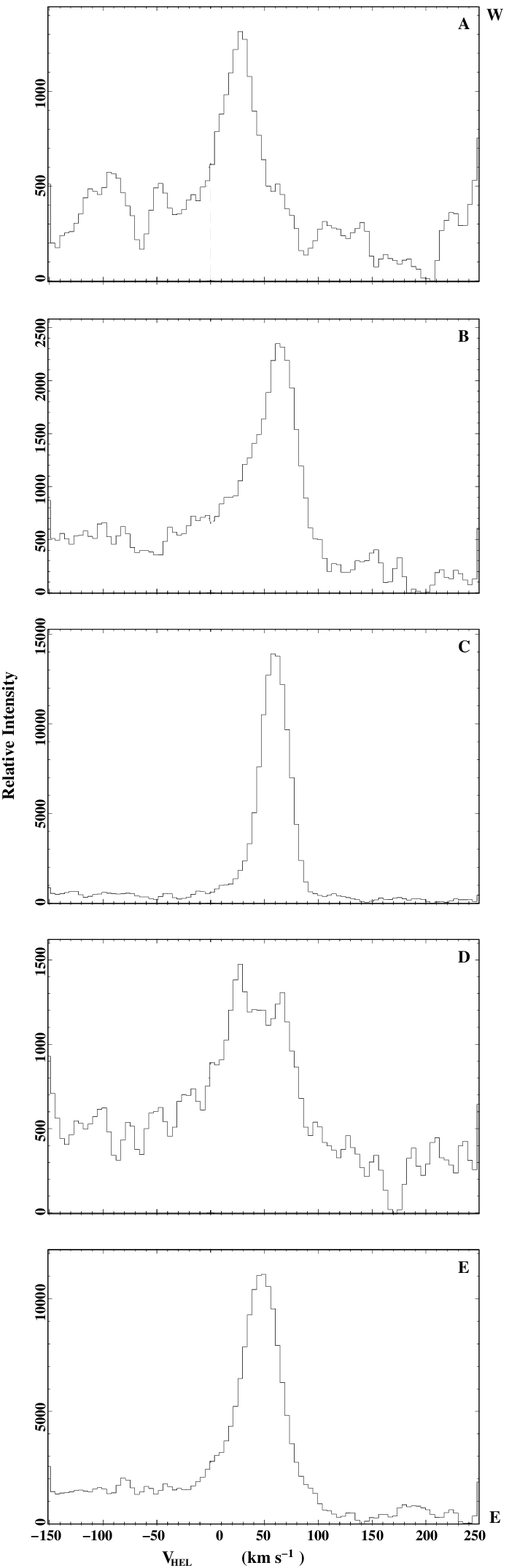}} 
\caption[]{The \nii\ line profiles from the incremental lengths marked
in Fig. 9.} 
\label{fig11} 
\end{figure*} 

\begin{figure*} 
\centering 
\scalebox{0.80}{\includegraphics{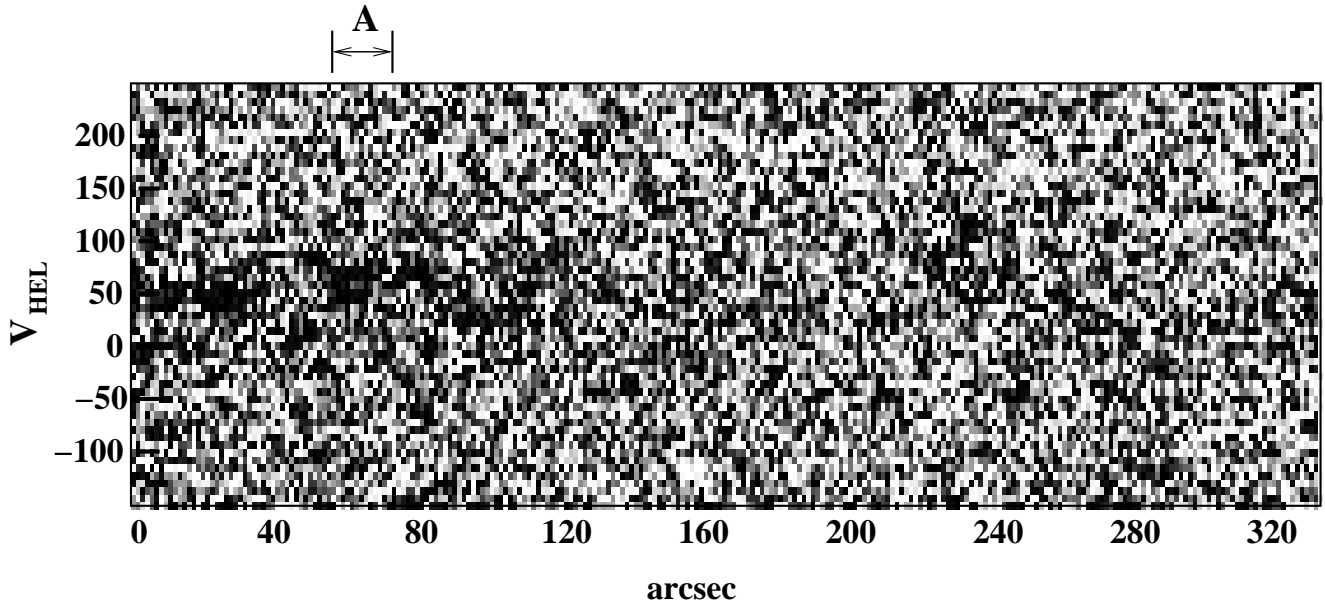}} 
\caption[]{A negative greyscale representation of the position
velocity array of \nii\ line profiles of the northern ridge.The slit
was orientated NS (see Fig. 6). The whole slit length is shown in
order to emphasize that detection of line profiles has only occurred
over a small part of the slit length i.e. we are not detecting diffuse
galactic emission along the same sight line. North is to the left.}
\label{fig12} 
\end{figure*}

\begin{figure*} 
\centering 
\scalebox{1.0}{\includegraphics{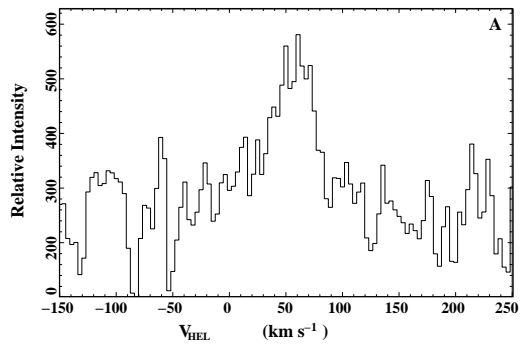}} 
\caption[]{The \nii\ line profile from the incremental length A marked
in Fig. 12.} 
\label{fig13} 
\end{figure*}

\bsp  
  
\label{lastpage}  
  
\end{document}